\documentclass[conference]{IEEEtran}
\IEEEoverridecommandlockouts
\usepackage{cite}
\usepackage{amsmath,amssymb,amsfonts}
\usepackage{algorithmic}
\usepackage{graphicx}
\usepackage{textcomp}
\usepackage{xcolor}
\usepackage{multirow}
\usepackage{hyperref}
\hypersetup{
    colorlinks=true,
    linkcolor=blue
    }

\def\BibTeX{{\rm B\kern-.05em{\sc i\kern-.025em b}\kern-.08em
    T\kern-.1667em\lower.7ex\hbox{E}\kern-.125emX}}
\begin{document}

\title{SOME/IP Intrusion Detection using Deep Learning-based Sequential Models in Automotive Ethernet Networks\\
}

\author{\IEEEauthorblockN{1\textsuperscript{st} Natasha Alkhatib}
\IEEEauthorblockA{\textit{LTCI} \\
\textit{Telecom Paris, IP Paris}\\
Palaiseau, France \\
natasha.alkhatib@telecom-paris.fr}
\and
\IEEEauthorblockN{2\textsuperscript{nd} Hadi Ghauch}
\IEEEauthorblockA{\textit{LTCI} \\
\textit{Telecom Paris, IP Paris}\\
Palaiseau, France \\
hadi.ghauch@telecom-paristech.fr}
\and
\IEEEauthorblockN{3\textsuperscript{rd} Jean-Luc Danger}
\IEEEauthorblockA{\textit{LTCI} \\
\textit{Telecom Paris, IP Paris}\\
Palaiseau, France \\
jean-luc.danger@telecom-paris.fr}
}

\maketitle
\begin{abstract}
Intrusion Detection Systems are widely used to detect cyberattacks, especially on protocols vulnerable to hacking attacks such as SOME/IP. In this paper, we present a deep learning-based sequential model for offline intrusion detection on SOME/IP application layer protocol. To assess our intrusion detection system, we have generated and labeled a dataset\footnote{Dataset URL:  \url{https://github.com/Alkhatibnatasha/SOMEIP_IDS}} with several classes representing realistic intrusions, and a normal class - a significant contribution due to the absence of such publicly available datasets. Furthermore, we also propose a recurrent neural network (RNN), as an instance of deep learning-based sequential model, that we apply to our generated dataset. The numerical results show that RNN excel at predicting in-vehicle intrusions, with F1 Scores and AUC values greater than 0.8 depending on each intrusion type. 
\end{abstract}


\section{Introduction}
Automobiles are no longer solely made up of mechanical systems. In fact, mechanical components have been taken over by electronics called ``Electronic Control Units" ECUs. These connected ECUs through various in-vehicle network infrastructures (CAN, FlexRay, MOST, and LIN) are in charge of making various car functions possible. However, these traditional in-vehicle networks have many limitations in terms of bandwidth and higher layer protocols. An adaptable and scalable in-vehicle network technology is thus required to realize sophisticated and innovative customer functions such as Adaptive cruise control, Collision avoidance, Driver drowsiness detection, Lane departure warning and others. To fulfill these automotive requirements, \textbf{Automotive Ethernet} technologies have been developed and standardized.

The deployment of Ethernet-based communication in in-vehicle network systems has several other benefits, such as the ability to reuse the associated OSI layers' protocols built and tested in other industries \cite{ae}. Furthermore, this cutting-edge technology enables the invention of new protocols for individual layers while reusing protocols for the rest such as the development of the automotive application layer protocol \textbf{Scalable service-Oriented Middle-warE over IP} (SOME/IP) \cite{someip_doc}.

SOME/IP is commonly used for relevant automotive applications due to its service-based communication approach and its adaptability to different automotive operating systems (e.g., QNX, OSEK and Linux) \cite{ae}. In other words, SOME/IP is increasingly adopted to coordinate the exchange of various services between disjoint applications on distinct ECUs. These services cover notifications about in-vehicle events, as well as  Remote Procedure Call (RPC) functions that enable an ECU client to request information from an ECU server. However, no security measures, such as authentication or encryption, are defined in the SOME/IP protocol specification \cite{someip}. In fact, the absence of SOME/IP security protocols may set the ground for an attacker to exploit a legitimate automotive system and initiate attacks from inside the network, such as intercepting and manipulating messages between two ECUs and other significant threats. To reduce the risk of the various inherent security threats, a robust defense plan is needed, which first requires detecting and analyzing these vulnerabilities. 

Due to their large approximation capacity, deep learning-based approaches are well-suited to detect network intrusions in various network types \cite{can} \cite{kang}. In this work, we have developed a deep learning based sequential model to detect network intrusions on the SOME/IP protocol. Sequential models are a category of deep learning model, where the training set is known (a-priori) to have a dominant temporal or causal  component: indeed, packets in a session of the SOME/IP protocol exhibit a strong temporal correlation, as each packet depends on previous ones. In the current work, we will contribute to the development of a sequence-based SOME/IP dataset, as no public SOME/IP dataset exists. Specifically, we generate and label a SOME/IP dataset, with four classes of general intrusion packets, as well as a class of normal packets. Moreover, our proposed deep learning-based model, a recurrent neural network (RNN) is able to classify these four intrusions on packets' sequences and the normal ones, with very large accuracy and F1 score.  Furthermore, we will evaluate our deep learning-based sequential model using the generated dataset.

Towards this end, our paper is organized into six sections. Section 2 discusses main publications that are related to SOME/IP intrusion detection. In section 3, we present an overview of the SOME/IP protocol. In section 4, we present our dataset and the different considered attacks. The suggested sequential model is presented in Section 5. In section 6, we present the different evaluation metrics used for performance evaluation.  We discuss our experimental results in section 7. Finally, we conclude our paper with future work direction.

\begin{figure}[h]
\centering
\includegraphics[scale=0.45]{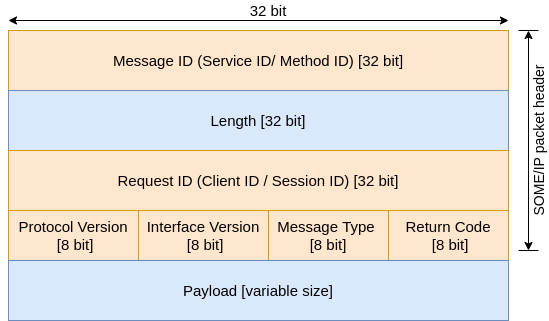}
\centering
\caption{SOME/IP packet}
\label{header}
\end{figure}

\section{Related Work}
Deep Learning approaches were highly used in previous works to detect network intrusions on the traditional in-vehicle network protocol \textbf{CAN}  \cite{gmiden} \cite{rehman} \cite{seo} \cite{kang}. However, no previous work has been addressed to detect intrusions on Automotive Ethernet especially SOME/IP protocol using Deep Learning due to the following reasons.
\begin{itemize}
    \item \textbf{Lack of Automotive Ethernet dataset: } The existence of large CAN databases \cite{can_data} containing both normal and abnormal network traffic behaviour has resulted in extensive research into deep learning applications on CAN. However, SOME/IP application layer protocol does not have well-known dataset available. Despite the fact that a new Automotive Ethernet dataset is recently published \cite{article_ae}, it is not helpful for our current work since it covers normal and abnormal streams of audio-video transport protocol (AVTP) which is different than SOME/IP protocol. Thus, the generation of the labeled dataset (and its publication) is one (but not the only) contribution of the current paper. 
    \item \textbf{Automotive Ethernet Standard gaining momentum: } Automotive Ethernet, a recent network protocol for vehicles, is gaining increasing momentum in standards for connected vehicles. 
\end{itemize}
\begin{figure}[h]
\centering
\includegraphics[scale=0.5]{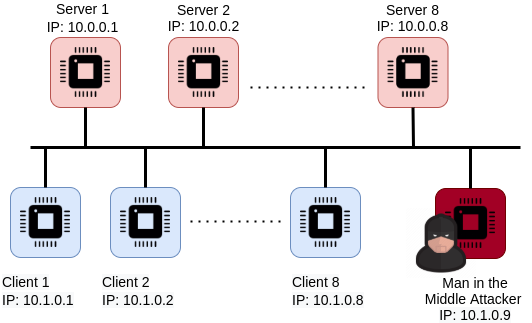}
\centering
\caption{Configured Network - Different SOME/IP clients and servers exchanging SOME/IP services over Automotive Ethernet Bus. Besides, a Client ECU is being compromised by an MITM Attacker.}
\label{fig:network}
\end{figure}
In terms of SOME/IP's latest security vulnerability investigations, researchers have begun to investigate its key vulnerabilities that could lead to cyberattacks on the in-vehicle network, as well as to develop IDS using different approaches. 
Gehrmann et al. \cite{someip_tobias} addressed specific problems and opportunities for intrusion detection in SOME/IP, as well as suggested an architecture for a SOME/IP intrusion detection scheme and discussed its security features. 
Iorio et al. \cite{someip_iorio} \cite{someip_iorio2} proposed a novel architecture to enhance the security of evolving SOME/IP middleware. 
Li at al. \cite{someip_li} developed Ori, a Greybox Fuzzer that can efficiently detect breaches in SOME/IP applications. 
Lauser et al. \cite{someip_lauser} have discussed
how formal models can be used to verify the security
of protocols used in modern vehicles.
Rumez et al. \cite{someip_rumez} explained various security countermeasures in the fields of firewalls, intrusion detection systems (IDSs), and identity and access management. 
Herold et al. \cite{someip} proposed a rule-based IDS for SOME/IP protocol.

To the best of our knowledge, none of the previous works have applied deep learning-based sequential models for intrusion detection on SOME/IP protocol. That is a main contribution of this work, in addition to the generation of the labeled  dataset (with multiple classes of intrusions and a normal class).

\begin{figure*}[h]
\centering
\includegraphics[scale=0.33]{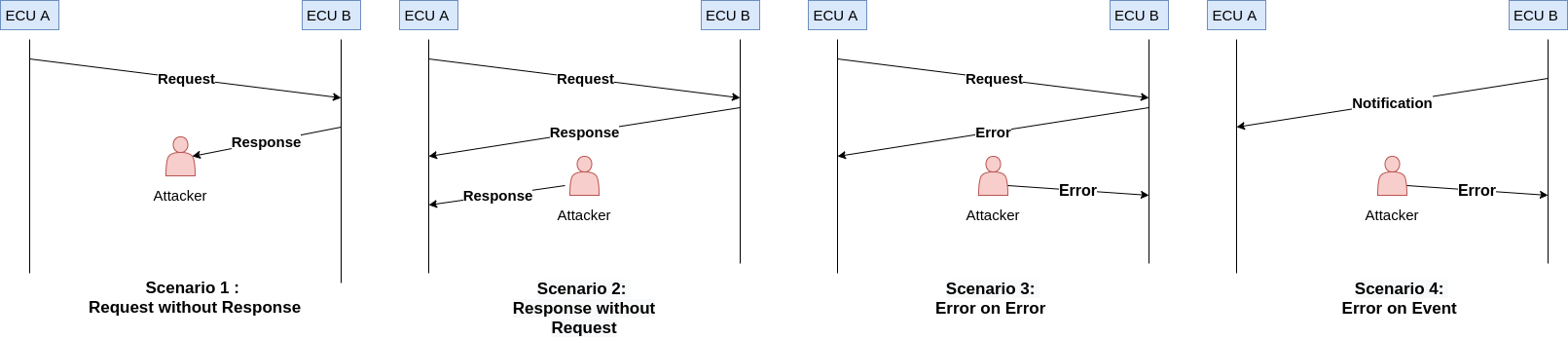}
\centering
\caption{Attacks on SOME/IP protocol}
\label{fig:scenarios}
\end{figure*}

\section{Overview of SOME/IP}
A \textbf{``Middleware"} refers to a connective tissue between different software applications. In other words, it handles all functions that are needed for a \textbf{``service"} to allow data exchange between several ECUs\cite{ae}. Due to the growing amount of software \cite{software} in automobiles, as well as the spread of functions within their in-vehicle network and the deployment of a variety of software architectures and operating systems inside vehicles, the implementation of a middleware software within in-vehicle networks is essential in bridging the gap between them.
Hence, after being proposed by the BMW Group in 2011 and standardized by AUTOSAR, SOME/IP was chosen as the standard middleware for IP-based service-oriented communication in cars. The middleware SOME/IP runs at top levels of the OSI model \cite{someip_tobias}. The structure of its header layout is as shown in Figure \ref{header}. Some of the fields presented in the header of the SOME/IP packet will be considered in our work as the input features for the deep learning based IDS (Table \ref{tab:features}).

\subsection{SOME/IP Remote Procedure Calls}
Since SOME/IP is a service-based communication approach, it allows the exchange of different types of remote procedure calls. In general, a remote procedure call RPC is an inter-process communication technique that is used for client-server based applications \cite{someip_doc}. In this current work, we have considered three main types of SOME/IP RPC :  
\begin{itemize}
    \item \textbf{Request/Response}: a method with Request and Response messages. The Request is a message sent by the client when it invokes a method. The Response is a message sent from the server to the client that contains the method invocation's outcome. 
    \item \textbf{Fire and Forget}: a procedure that only uses Request messages. As in the Request/Response scenario, the client calls a server method. However, unlike in the Request/Response instance, the client does not anticipate a response.
    \item \textbf{Events}: In this method, the server sends messages to the client with particular information either periodically or whenever there is a change (event). The server expects no response from the client. 
\end{itemize} .

\begin{table}[h]
\begin{center}
\caption{Training and Testing Dataset classes}
\begin{tabular}{||c | c |c ||} 
\hline
\textbf{Class} & \textbf{Training Dataset}& \textbf{Testing Dataset}\\
\hline\hline
Normal &  2533 &  2471\\ 
\hline
Error on Error & 39 &  54\\
\hline
Error on Event & 60 &  54\\
\hline
Missing Response & 92 &  81\\
\hline
Missing Request & 83 &  111\\
\hline
\hline 
\textbf{Total} & \textbf{2807}& \textbf{2771}\\
\hline
\end{tabular}
\label{tab:classes}
\end{center}
\end{table}

\section{Generating Labeled Dataset}

\subsection{Dataset Generation}
\subsubsection{SOME/IP packet generator}
\begin{figure}[h]
\centering
\includegraphics[scale=0.3]{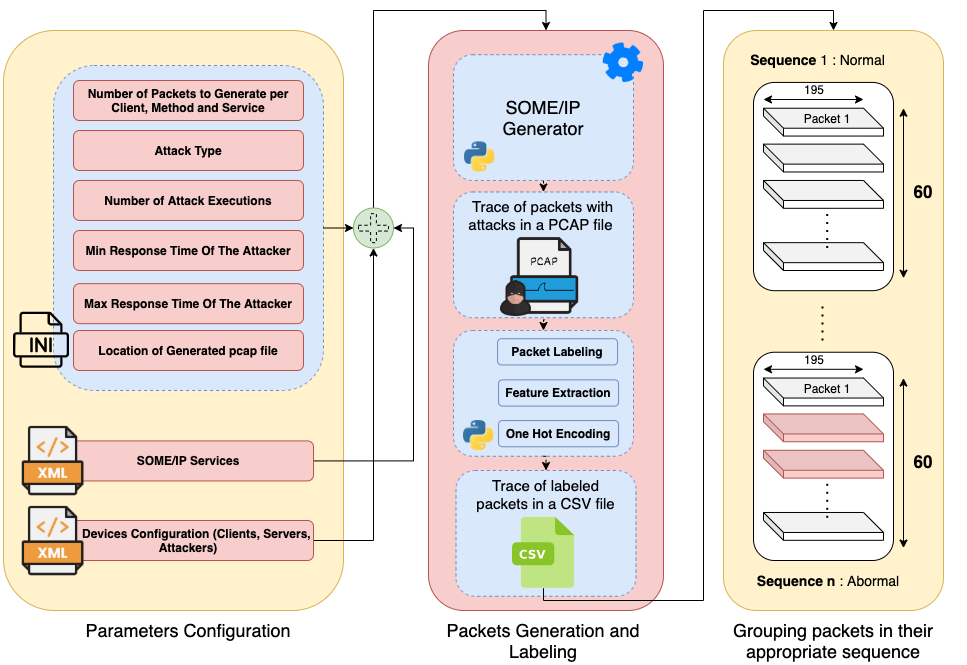}
\centering
\caption{Dataset Generation}
\label{fig:dataset}
\end{figure}

In order to generate SOME/IP libpcap dump files, we have used the SOME/IP Generator developed by \cite{someip}, implemented in Python 3 and available in Github \cite{someip_generator}. The generator models the behavior of different clients and servers assumed to behave according to the AUTOSAR standard specification, as well as an attacker carrying out a variety of attacks depicted in Figure \ref{fig:scenarios} and described in section \ref{someip_attacks}. As seen in Figure \ref{fig:dataset} and Table \ref{tab:parameters}, we have tuned the different parameters for generating different attack scenarios such as the network architecture configuration depicted in Figure \ref{fig:network}, the SOME/IP services to be exchanged, and the attack to be implemented along with its frequency of execution. For training and testing our deep learning based IDS, we have generated several pcap files corresponding to different attack types, concatenated them and processed them as described in section \ref{process}. The distributions of both datasets are shown in Table \ref{tab:classes}, and their corresponding features are described in Table \ref{tab:features}. The training dataset comprises about 274 attacks, is 132 MB in size, and contains 2807 packets. Regarding the testing dataset, it contains around 300 attacks, has a size of 130 MB and composed of 2771 packets. Readers can get our SOME/IP intrusion dataset by referring to \cite{natasha}.

\begin{table*}[h]
\begin{center}
\caption{Tuned Parameters for Dataset Generation (represents first step in Figure \ref{fig:dataset})}
\begin{tabular}{|l | l| l|} 
\hline
\textbf{Parameters to configure} & \textbf{Description} & \textbf{ Chosen Value}\\
\hline\hline
\multirow{2}{*}{Devices} & Contains information like name, type, mac, ip & 8 Servers, 8 Clients and  \\
                         & sender port and receiving port of each Client, Server and Attacker & 1 Attacker\\
\hline
Services & Contains information about offered and requested services & 3 Services\\
\hline
\multirow{1}{*}{Number of packets to generate per Client, Method and Service} & Defines the number of packets generated per client & 50\\
\hline
Number of attacks to execute & Defines the rate an attack will be performed & 10\\
\hline
Minimum Response Time of Attacker & Defines the minimum response time of the attacker in ms & 1\\
\hline
Maximum Response Time of Attacker & Defines the maximum response time of the attacker in ms & 3\\
\hline
\multirow{4}{*}{Implemented Attack}& & Error On Error\\      
                                   & Defines which attacks can be used &  Error On Event \\
                                   & & Missing Request \\
                                   & & Missing Response\\
\hline
Output File Location & Describes the location where to store the resulting pcap & output.pcap \\
\hline
\end{tabular}
\label{tab:parameters}
\end{center}
\end{table*}

\subsubsection{Dataset Imbalance}
As seen in table \ref{tab:classes}, the distribution of samples across the different classes is biased. In fact, the attack classes frequency is highly imbalanced, i.e., there is a bias or skewness towards the majority class (Normal class) present in the target. It is reasonable to have such a skewed dataset since it represents an anomaly problem. However, we do not aim changing the nature of the data and make it balanced even though this problem poses a challenge for predictive modeling as most of the supervised deep learning algorithms used for classification were designed around the assumption of an equal number of examples for each class. An alternative solution would be the adoption of specialized techniques such as \textbf{Adaptive Weighting} \cite{weights}. This technique, implemented in our work, is considered as a popular approach for imbalance learning since it weighs samples in rare classes with high cost and then applies cost-sensitive learning methods to deal with imbalance in classes. Table \ref{tab:class_weight} presents the weights assigned for each class, for the dataset considered in this work.

\begin{table}[h]
\begin{center}

\caption{Class Weight}
\begin{tabular}{||c | c ||} 
\hline
\textbf{Class} & \textbf{Class Weight}\\
\hline\hline
Normal &  0.16 \\
\hline
Error on Event & 6.68 \\
\hline
Error on Error & 10.35 \\
\hline
Missing Response & 4.33 \\
\hline
Missing Request & 4.86 \\
\hline

\end{tabular}
\label{tab:class_weight}
\end{center}
\end{table}

\begin{table}[t]
\begin{center}
\caption{Dataset Features}
\begin{tabular}{|l | l |} 
\hline
\textbf{Feature} &
\textbf{Description}\\
\hline\hline
Service ID & A unique identifier for a service  \\ 
\hline
\multirow{2}{*}{Method ID }& A unique identifier of a method,  \\
                           & an event or a field that belong to the service\\ 
\hline

\multirow{2}{*} {Client ID} &  Allows a server to differentiate \\
                            & calls from multiple clients to the same method \\ 
\hline
\multirow{3}{*} {Message Type} & Used to differentiate different types of                                      messages \\ 
                               & such as : request,request no return, \\
                               & notification, response and error \\ 
\hline
\multirow{2}{*}{Session Id} & Allows a subscriber to differentiate multiple\\                               &  calls to the same method \\ 
\hline
Interface Version & Contains the
Major Version of the Service Interface \\ 
\hline
Protocol Version & Contains the
SOME/IP protocol version \\ 
\hline
\multirow{2}{*} {Return Code} &  Used to signal whether a request\\
                            & was successfully processed \\ 
\hline
IP source & IP of the sending device \\ 
\hline
IP destination & IP of the receiving device\\ 
\hline
Protocol & Application layer protocol\\ 
\hline
Source Port & Port number of the sending device\\ 
\hline
Destination Port & Port number of the receiving device\\ 
\hline
Mac source  & MAC Address of the sending device  \\ 
\hline
Mac destination  & MAC Address of the receiving device\\ 
\hline
\multirow{3}{*} {Label} & Specifies the class of each packet such as normal,\\                         & error on error, error on event, request without \\                           &  response, response without request\\ 
\hline
\end{tabular}
\label{tab:features}
\end{center}
\end{table}

\subsubsection{SOME/IP intrusions}
The SOME/IP attacker is able to compromise a known device within the system. Thus, it has a valid MAC address, IP address, and service ID. It eavesdrops on all traffic within the network and send packets to all clients and all servers, and thus impersonates other SOME/IP devices and services \cite{someip}. Through our work, we are interested in cyberattacks which lead to deviations from the protocol specifications in a communication session between two devices (as seen in Figure \ref{fig:scenarios}) and which can be detected using deep learning based sequential Models. These four intrusion types considered in this work are detailed below (illustrated also in Figure \ref{fig:scenarios}): 
\label{someip_attacks}
\begin{itemize}
    \item \textbf{Requests without Response}: Requests have to be answered with either a response or an error message. If a request was never answered, it means that an attacker has relayed the communication between the client and the server who believe that they are directly communicating with each other.
    \item \textbf{Response without Request}: A response should only be delivered in response to an open, previous request. As a result, a normal request with message type 0x00 should be answered by a single response with message type 0x80. Two replies to a single request break the protocol and may indicate the existence of an attacker attempting to impersonate the server and injecting extra packets.
    \item \textbf{Error on Error}: Based on AUTOSAR standard specification, an error message should not be answered with another error message. Hence an incoming error which doesn't have a corresponding request (or other packet) with the same settings indicates the presence of a network intrusion. 
    \item \textbf{Error on Event}: Notifications should not be answered with an error message. Thus, a notification replied to with an error depicts a network intrusion between the client and the server. 
\end{itemize}

\subsubsection{Data Preparation}
\label{process}
This section describes the different steps (seen in Figure \ref{fig:dataset}) achieved for our dataset to be fed to our deep learning based IDS for training and testing. 
\begin{enumerate}
    \item \textbf{Packets generation and labeling: }
    The SOME/IP packet generator is able to generate pcap files composed of unlabeled packets gathered from the whole network. Since we are using a supervised learning approach, we had to label each packet. Hence, a packet is labeled by 0 if it behaves according to the AUTOSAR standard specification. Otherwise, it is labeled by 1,2,3 or 4 if it represents error on event, error on error, request without response or response without request attacks respectively.
    \item \textbf{Packets Feature Extraction and One-Hot Encoding: }
    Each packet is represented by 16 categorical features, described in Table \ref{tab:features}. However, these features had to be converted to binary vectors using one-hot encoding technique. In fact, many deep learning algorithms cannot work with categorical data directly. Hence, the categories must be converted into numbers. This is required for both input and output variables that are categorical. After encoding, the 15 features that represented input variables were extended to 195 features and the output variable (Label) was extended to 5 classes. 
    \item \textbf{Grouping Packets into Sequences: }
    In order to detect intrusions affecting the communication behavior between two devices, we had to group packets that belong to each communication in their appropriate sequence. Hence, each sequence represents a series of ordered packets exchanged between a client and a server with the same session identifier. As seen in Figure \ref{fig:dataset}, we have grouped packets in their corresponding sequences. Thus, our IDS will detect the presence of an intrusion in a communication between a client and a server by inspecting each sequence of packets.  However, since we are dealing with variable length sequence prediction problems, our data had be transformed such that each sequence has the same length. Hence, after transformation, each sequence contains 60 packets which is the maximum number of packets per sequence, i.e, a sequence is padded by zeros if it contains less than 60 packets per session. Our dataset was generated with the constraints that only one type of attack can occur between two devices. Therefore, sequences are either labeled as normal or by a number corresponding to only one of the (four) possible intrusions. Furthermore, an attack begins and ends in the same session between a client and a server. Hence, an attack cannot be executed in different sessions at the same time. 
    \item \textbf{Sequences Concatenation: }
    Finally, after labeling the different sequences that represent diverse attacks, we have concatenated them in a single dataset that will used for training and testing the deep learning based IDS. 
\end{enumerate}

\section{Proposed Sequential Model}
\begin{figure}[h]
\centering
\includegraphics[scale=0.4]{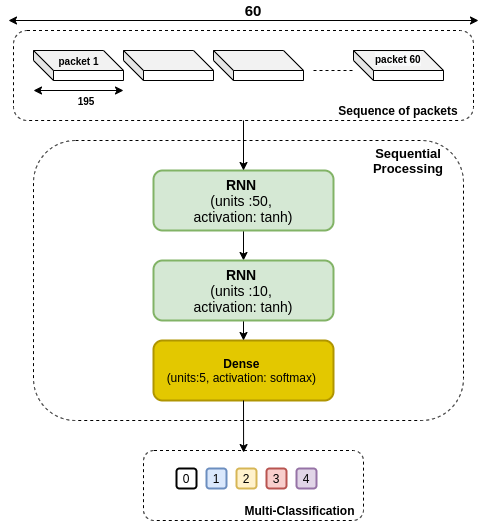}
\caption{RNN based IDS architecture}
\label{fig:architecture}
\centering
\end{figure}

Deep Learning based sequential models have been widely adopted to detect intrusions and anomalies in various type of computer networks \cite{seq1} \cite{seq2} \cite{seq3}. Our proposal in this section is to employ Recurrent Neural Networks or RNN as a sequential model to the labeled dataset generated in the previous section. The proposed RNN is presented in Figure \ref{fig:architecture}. Furthermore, its resulting hyperparameters are shown in Table \ref{tab:hyperparameters}. The input to the RNN consists of 60 ordered packets with 195 features each. It passes to two stacked RNN layers which have recurrent connections between hidden units. The two RNN layers read the entire input sequence of 60 packets and feed their output to a dense layer which produces 5 outputs (corresponding to each the 5 classes) using softmax function. We denote the training set, $ \{ ( x_t , y_t ) \}_{ t = 1}^T  $, where $x_t $ is the feature vector (a vector of dimension $195$) for sample $t \in \{ 1, ..., T \} $, $ y_t $ represents the corresponding label for sample $t \in \{ 1, ..., T \} $, and $T$ the number of samples in the training set.  Moreover,  each label in the training set is such that, $y_t$ is a binary vector of dimension $5$, i.e., $y_t \in \mathbb{B}^5 $, where  element $i \in \{1, ..., 5  \}$ of the vector  $y_t$ is a binary variable representing whether the corresponding feature vector, $x_t$, belongs, i.e., corresp entry $=1$ (or does not belong, i.e., corres entry $=0$ ) to class $i \in \{1, ..., 5  \}$. Furthermore, $y_t$ may only have one non-zero entry, which follows from our previous assumption that only one intrusion is possible in each sample.  

The equations describing the operation for the RNN are the following : 
\begin{equation}
  a_{t} = W h_{t-1} + U x_{t} + b, \forall t \in \{ 1, ..., T-1 \}.
\end{equation}
\begin{equation}
  h_{t} = \psi (a_{t})
\end{equation}   
\begin{equation}
  o_{t} = V h_{t} + c
\end{equation}
\begin{equation}
  \tilde{y_{t}} = \phi(o_{t})
\end{equation}

where the vector $a_{t}$ is a linear combination between $x_{t}$, the feature vector for sample $t$, and the hidden layer output of the RNN for sample $t-1$, $h_{t-1}$. $h_{t} $ is a vector modeling the hidden layer output of the RNN for sample $t$. $o_{t}$ (a vector of dimension $5$) is a linear combination of the output hidden layer $h_{t}$. $\tilde{y_{t}}$ is the prediction that RNN outputs for sample $x_{t}$ , and has the same properties as $y_t$. $W$, $U$, $V$, are the shared weights matrices that will optimized in training. $\psi$ and $\phi$ are non-linear activation functions, applied element-by-element on their respective inputs.

\begin{table}[h]
\begin{center}
\caption{Hyperparameters}
\begin{tabular}{||c | c | c |c ||} 
\hline
\textbf{Hyperparameters} & \textbf{Values}\\
\hline\hline
\multirow{ 1}{*}{Number of layers} & 3\\ 
\hline
\multirow{ 1}{*}{Number of Neurons per layer} & (50,10,5)\\ 
\hline
\multirow{ 1}{*}{Activation Function per layer} & (tanh,tanh,softmax)\\ 
\hline
\multirow{ 1}{*}{Optimizer} & Adam \\ 
\hline
\multirow{ 1}{*}{Loss} & Categorical Cross Entropy \\ 
\hline
\multirow{ 1}{*}{Learning Rate} & 0.001\\ 
\hline
\multirow{ 1}{*}{Batch size} & 100\\ 
\hline
\multirow{ 1}{*}{Epoch size} & 50\\ 
\hline
\end{tabular}
\label{tab:hyperparameters}
\end{center}
\end{table}

\section{Evaluation Metrics}
We use the Area Under The Curve (AUC) values, Receiver Operating Characteristics (ROC) curves and F1 scores and calculate them for each class to assess our IDS performance. 

We also present the multi-class confusion  matrices, which contains information about the actual and prediction classifications done by the classifier, to describe the performance of the multi-classifier models. The training samples corresponding to the  label (ground truth) $y_t$,  are represented by each row of the matrix, whereas the occurrences in a predicted label $\hat{y_t}$ (RNN output), are represented by each column. Specifically, for the task at hand, the confusion matrix will be a $5 \times 5 $,  where element $(i,j) \in \{1,...,5 \} \times \{1,...,5 \}$ denotes the normalized number of occurrences , where the true label is from class $i \in \{ 1,..., 5 \} $, and the predicted label is from class $ j \in \{ 1, ..., 5 \}$. Thus, for an ideal multi-class classifier all the diagonal entries should be $1$, while the off-diagonal entries should be $0$. 

In addition to the confusion matrix, we use the following other metrics. 

Recall is the ratio of correctly predicted
positive observations of all the observations in the actual class.
\begin{equation}
Recall =\frac{TP}{TP + FN}
\end{equation}
Precision is the ratio of correctly predicted
positive observations of all the observations in the predicted class.
\begin{equation}
Precision =\frac{TP}{TP + FP}
\end{equation}
Hence, the F1-Score is calculated using the following equation: 
\begin{equation}
F1-Score =\frac{2 \times Precision \times Recall}{Precision + Recall}
\end{equation}

Where: TP = True Positive; FP = False Positive; TN = True
Negative; FN = False Negative.

Since the dataset is imbalanced, the F1 score which is the weighted average results of both metrics precision and recall is  essential for evaluating the deep learning based IDS performance. The model has a large predictive power if the F1 score is near to 1.0.

The receiver operating characteristic curve, or ROC curve, is a graphical representation of a classifier system's performance while its discrimination threshold is modified. It is calculated by displaying the true positive rate (TPR) vs the false positive rate (FPR) at different threshold levels. The ideal classifier should provide a point in the upper left corner of the ROC space, or coordinate (0,1), signifying 100 percent sensitivity (no false negatives) and 100 percent specificity (no false positives). AUC stands for "Area under the ROC Curve." It measures the entire two-dimensional area underneath the entire ROC curve. The AUC for an ideal classifier should be $1$.

\begin{figure*}[t]
\centering
\includegraphics[scale=0.5]{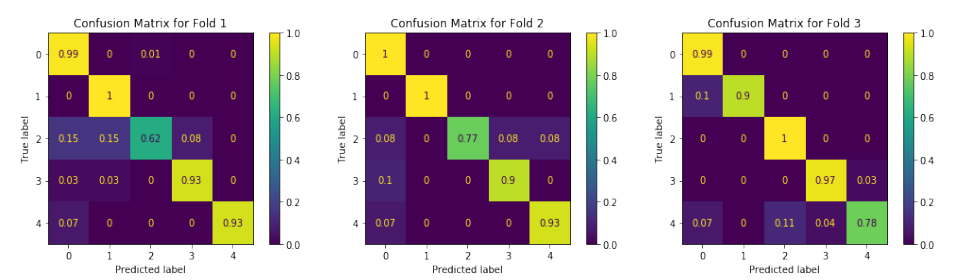}
\caption{Confusion matrices for three different models on Validation dataset}
\centering
\label{fig:cm_validation}
\end{figure*}

\begin{figure*}[t]
\centering
\includegraphics[scale=0.5]{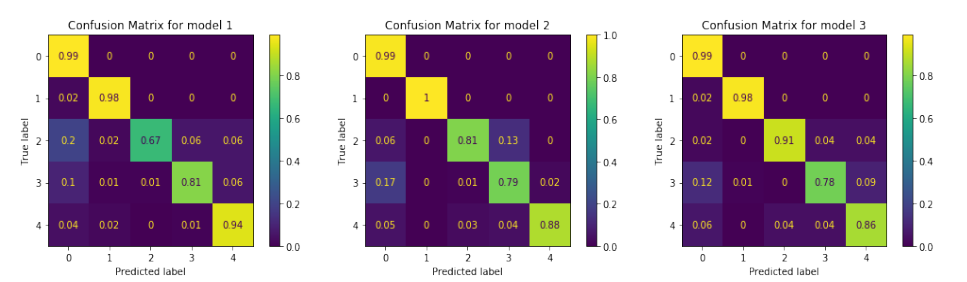}
\caption{Confusion matrices for three different models on Testing dataset}
\centering
\label{fig:cm_testing}
\end{figure*}

\begin{figure*}[t]
\centering
\includegraphics[scale=0.5]{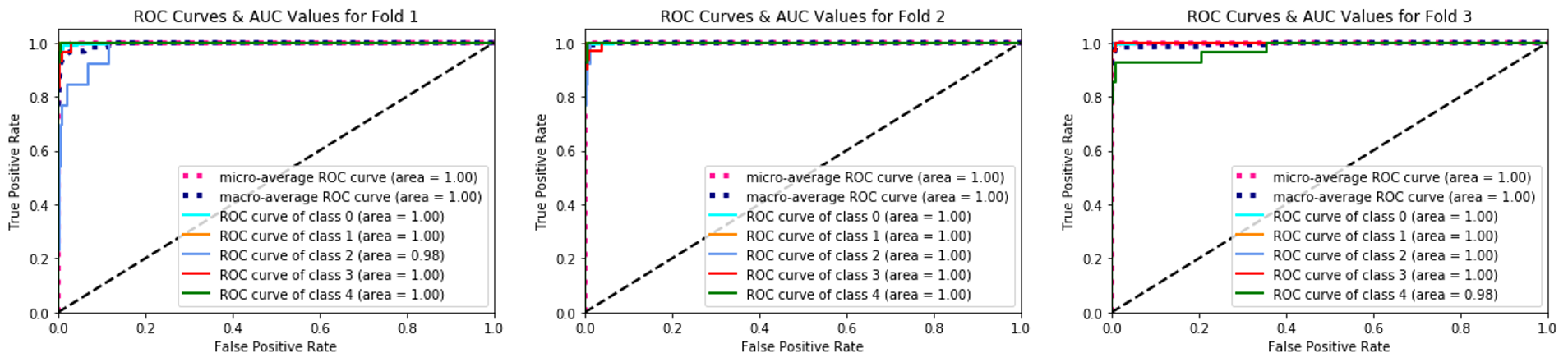}
\caption{ROC Curves and AUC values of each class of the Validation datasets}
\centering
\label{fig:roc_validation}
\end{figure*}

\begin{figure*}[t]
\centering
\includegraphics[scale=0.5]{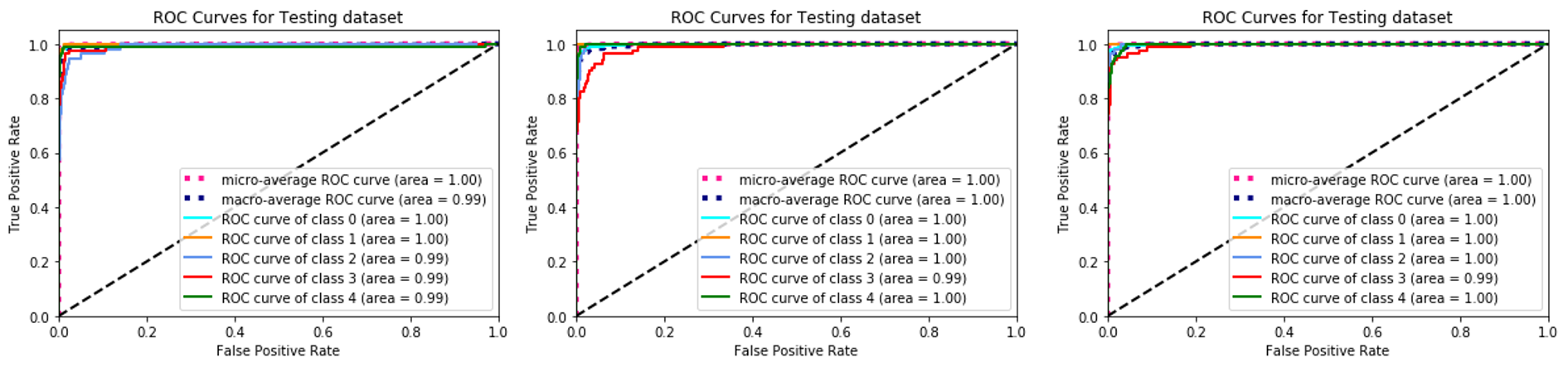}
\caption{ROC Curves and AUC values of each class of the Testing dataset}
\centering
\label{fig:roc_testing}
\end{figure*}
In the experiments, we use the Python library Keras \cite{keras} to implement our RNN model. We train and evaluate our model on an Intel(R) Core(TM) i5-6440HQ CPU @ 2.60GHz.

\section{Results}

Using the generated dataset, we ran a three-fold cross-validation with early stopping to ensure a large statistical confidence for our model's prediction performance. In each cross-validation, 67\% and 33\% of the data are chosen at random as the training and validation sets, respectively. The training set is used for model fitting and the validation set is used for model evaluation for each of the hyperparameter sets. Furthermore, they have the same proportion of classes in each validation fold. After cross-validation, we got three trained RNN models. To assess the overall performance of our approach, we ran three experiments on the testing dataset, one for each trained model.

\begin{table}[h]
\begin{center}

\caption{Results on Validation Data}
\begin{tabular}{||c | c | c |c |c ||} 
\hline
\textbf{Fold} & \textbf{Class} & \textbf{Recall} & \textbf{Precision} & \textbf{F1-Score} \\
\hline\hline
\multirow{ 5}{*}{1} & Normal &  0.99 & 0.99 & 0.99\\ 

& Error on Event &  1 & 0.87 & 0.93\\ 

& Error on Error &  0.61 & 0.61 & 0.61\\ 

& Missing Response &  0.93 & 0.93 & 0.93\\ 

& Missing Request &  0.93 & 1 & 0.96\\ 

\hline

\multirow{ 5}{*}{2} & Normal &  0.99 & 0.99 & 0.99\\ 

& Error on Event &  1 & 0.95 & 0.97\\ 

& Error on Error &  0.77 & 0.91 & 0.83\\ 

& Missing Response &  0.90 & 0.93 & 0.91\\ 

& Missing Request &  0.93 & 0.96 & 0.95\\ 

\hline

\multirow{ 5}{*}{3} & Normal &  0.99 & 0.99 & 0.99\\ 

& Error on Event &  0.9 & 1 & 0.95\\ 

& Error on Error &  1 & 0.87 & 0.93\\ 

& Missing Response &  0.97 & 0.88 & 0.93\\ 

& Missing Request &  0.77 & 0.87 & 0.82\\ 

\hline
\end{tabular}
\label{tab:results_validation}
\end{center}
\end{table}

\begin{table}[h]
\begin{center}
\caption{Results on Testing Data}
\begin{tabular}{||c | c | c |c |c ||} 
\hline
\textbf{Model} & \textbf{Class} & \textbf{Recall} & \textbf{Precision} & \textbf{F1-Score} \\
\hline\hline
\multirow{ 5}{*}{1} & Normal &  0.99 & 0.99 & 0.99\\ 

& Error on Event &  0.98 & 0.93 & 0.95\\ 

& Error on Error & 0.67 & 0.97& 0.79\\ 

& Missing Response &  0.81 & 0.88 & 0.84\\ 

& Missing Request &  0.94 & 0.93 & 0.93\\ 

\hline

\multirow{ 5}{*}{2} & Normal &  0.99 & 0.99 & 0.99\\ 

& Error on Event &  1 & 0.93 & 0.96\\ 

& Error on Error &  0.81 & 0.81 & 0.81\\ 

& Missing Response &  0.79 & 0.79 & 0.79\\ 

& Missing Request &  0.88 & 0.96 & 0.92\\ 

\hline

\multirow{ 5}{*}{3} & Normal & 0.99 & 0.99 & 0.99\\ 

& Error on Event & 0.98 & 0.98 & 0.98\\ 

& Error on Error & 0.91 & 0.82 & 0.86\\ 

& Missing Response &  0.78 & 0.84 & 0.81\\ 

& Missing Request & 0.86 & 0.89 & 0.87\\ 

\hline
\end{tabular}
\label{tab:results_test}
\end{center}
\end{table}

The classification results for three-fold cross-validation are shown in Table \ref{tab:results_validation}. The experimental results demonstrated that the model performed well, with acceptable F1-score values for each class of the validation folds. Thus, the models can classify almost all type of attacks on sequences correctly. Moreover, no significant difference in performance metrics exists across the three cross-validations. As a result, the training process is robust with the selected hyperparameters. 

Figure \ref{fig:cm_validation} show a summary of prediction results on the 3 folds for the multi-classification intrusion detection problem during cross-validation. The models have few prediction errors as values outside the diagonal of the confusion matrices approach zero. Hence, the models have well performed since most of the samples are located in the diagonal of the confusion matrices. 

Figure \ref{fig:roc_validation} presents the ROC curves and AUC values for each attack type and for each model (micro-ROC and macro-ROC curves) during the three-fold cross-validation. The displayed figures has AUC values near 1 which means the 3 models have a good measure of separability for the different attack types. Furthermore, the different roc curves have a point in the upper left corner or coordinate (0,1) of the ROC space for each model, representing the ability of the model to have a a huge sensitivity (no false negatives) and an outstanding specificity (no false positives).
We have then performed three tests using the three models that were trained during the cross-validation on the testing dataset. Based on the results shown in Table \ref{tab:results_test}, Figure \ref{fig:cm_testing} and \ref{fig:roc_testing}, we found that the overall performance of the models is outstanding. In fact, the trained models were  able to generalize to data that they haven’t seen before and did not merely learn to model the training data.
In average, the model has well predicted the normal behavior of packets in a sequence (F1-score =0.99). It is also able to predict the several other types of attack since F1-score value varies between 0.8 and 0.96. Moreover, the models outstanding performance is depicted in Figure \ref{fig:roc_testing} since the ROC curves of each class are closer to the top-left corner and the AUC values for the different classes approach 1.

\section{Limitations}
Our main contribution in this paper is to prove that deep learning algorithms are suitable for detecting intrusions on SOME/IP protocol and classifying them. Hence, our developed intrusion detection system is considered as an attack classifier rather than an anomaly detector. However, for future work, we aim to develop an anomaly based IDS able to detect any type of abnormal behavior and which can be trained using unsupervised learning. Furthermore, the dataset used for developing the proposed IDS is synthetic and has been processed for offline intrusion detection, i.e, an intrusion is detected when the session between a client and a server ends. However, for future work, we will be developing a intrusion detection system used for real-time detection and trained using an extracted dataset from real vehicle. 
\section{Conclusion}
SOME/IP is an automotive/embedded communication protocol which enhances intercommunication between several ECUs. In this paper, we have proposed a deep learning based IDS that can be leveraged to detect intrusions on SOME/IP automotive protocol. We have generated a labeled dataset to train and evaluate the performance of our model in offline mode and made it public for reproducibility. Performance results show that the proposed models can be successfully implemented to detect multiple types of intrusions on SOME/IP protocol, with  very large F1-Scores and AUC values bigger than 0.8 for each class. For future work, we aim to create a SOME/IP packet-based dataset extracted from a real vehicle and test diverse unsupervised learning based IDS that can be deployed for detecting unknown intrusions in real-time.

\vspace{12pt}

\end{document}